\documentclass{PoS}
\usepackage{amsmath}
\usepackage{graphicx}

\newlength{\dummysp}
\newcommand{\beq}{\begin{eqnarray}}
\newcommand{\eeq}{\end{eqnarray}}
\newcommand{\nnn}{ \nonumber \\ }

\newcommand{\ord}[1]{{{\cal O}(#1)}}
\newcommand{\gappeq}{\mathrel{\rlap {\raise.5ex\hbox{$>$}}
{\lower.5ex\hbox{$\sim$}}}}
\newcommand{\lappeq}{\mathrel{\rlap{\raise.5ex\hbox{$<$}}
{\lower.5ex\hbox{$\sim$}}}}

\newcommand{\ben}{\begin{enumerate}}
\newcommand{\een}{\end{enumerate}}

\newcommand{\ddd}{\nnn &&}

\newcommand{\bit}{\begin{itemize}}
\newcommand{\eit}{\end{itemize}}

\newcommand{\Abar}{{\bar A}}
\newcommand{\Bbar}{{\bar B}}
\newcommand{\Cbar}{{\bar C}}
\newcommand{\Xbar}{{\bar X}}

\def\[{\left [}
\def\]{\right ]}
\def\({\left (}
\def\){\right )}

\title{On the decoupling of mirror fermions}

\ShortTitle{Decoupling mirror fermions}

\author{\speaker{Joel Giedt} \\
Department of Physics, Applied Physics, and Astronomy, \\ 
Rensselaer Polytechnic Institute, 110 8th St., Troy, New York, 12180, USA \\
E-mail: \email{giedtj@rpi.edu}}

\author{Chen Chen}

\author{ Erich Poppitz \\
Department of Physics,   University of Toronto, 
Toronto, ON M5S 1A7, Canada \\
E-mail:  \email{poppitz@physics.utoronto.ca}}

\abstract{
We study an approach to chiral gauge theories on the lattice that involves
decoupling ``mirror'' fermions from a vector-like theory. 
We have computed the polarization tensor in the ``3-4-5'' theory and find a 
directional discontinuity that appears to be nonzero in the continuum limit. 
This strongly suggests that the mirror fermions do not decouple.
}

\FullConference{31st International Symposium on Lattice Field Theory - LATTICE 2013\\
		July 29 - August 3, 2013\\
		Mainz, Germany}

\begin{document}
  
\section{Introduction}
Strongly-coupled chiral gauge theories are prominent
in ideas about what lies beyond the Standard Model. 
Most attractive channel arguments suggest that they may
lead to tumbling dynamics \cite{Raby:1979my}, explaining
a hierarchy of scales in extended technicolor.  In supersymmetric models,
strong chiral dynamics is often a key component in models of dynamical
supersymmetry breaking (see,~e.g.,~the review \cite{Poppitz:1998vd} 
for extensive references).  
For these and other more theoretical reasons, we would like
to study strongly-coupled chiral gauge theories from first principles,
using a lattice formulation.  In contrast to vector-like theories,
severe difficulties arise when we try to formulate 
chiral gauge theories on the lattice.  In older approaches,
solving the fermion doubling problem resulted in actions that
explicitly violated chiral symmetry \cite{Nielsen:1980rz,Nielsen:1981hk,Nielsen:1981xu};
e.g., Wilson's fermion discretization \cite{Karsten:1980wd,Karsten:1981gd}.  
More recently, discretizations that
satisfy the Ginsparg-Wilson relation \cite{Ginsparg:1981bj}
have been developed \cite{Kaplan:1992bt,Narayanan:1993ss, Narayanan:1994gw,Neuberger:1997fp}, 
maintaining lattice chiral symmetry \cite{Luscher:1998pqa}.  
In spite of this progress, problems surfaced with this approach as it applies
to chiral gauge theories.  For instance, for overlap fermions the fermion measure
had a gauge-background dependent phase.  

Various attempts have been made to overcome the problems.  For instance,
the recent approach of \cite{Golterman:2004qv} involves non-perturbative gauge fixing.
By contrast, L\"uscher's approach defines the phase of the fermion determinant
through consistency conditions (see the review \cite{Luscher:2000hn}). 
In the case of Abelian chiral gauge theories, the phase can be 
determined for arbitrary gauge backgrounds though to date no
numerical implementation of this has been performed.
The approach studied in this paper would avoid having to 
solve L\"uscher's consistency conditions.  This ``mirror fermion decoupling''
has been reviewed in \cite{Poppitz:2010at}.

In this approach a vector-like theory is decomposed into
``light'' particles of one chirality and ``mirror''
particles of the opposite chirality.  
Through a strong Yukawa coupling of the mirror fermions to a unitary ``Higgs'' field it is
hoped that they will acquire a mass of order the ultraviolet cutoff.
Note that this occurs in the symmetric phase of the Higgs theory,
i.e., there is no spontaneous gauge symmetry breaking in our model.
The large mass is purely a dynamical effect of the strong
Yukawa coupling.  This is similar to \cite{Eichten:1985ft}, which used 
four-fermion interactions instead of a Higgs field.

Earlier efforts \cite{Eichten:1985ft, Smit:1985nu} did not have a clean
separation of chiral components of the vector-like theory.
Thus both the light and mirror sectors experienced 
the strong interactions.  The result was that the spectrum did not
separate according to the chiral goals.  This situation
has changed because of the Neuberger-Dirac overlap operator and exact lattice chirality.
Using this, there was a proposal to lift the mirror sector with
a strong Yukawa coupling that does not affect the light sector
\cite{Bhattacharya:2006dc} (note also the earlier proposal of Ref.~\cite{Creutz:1996xc}
using domain wall fermions).
 
An initial numerical study of this proposal was performed in Ref.~\cite{Giedt:2007qg},
using as an example the two-dimensional Schwinger model, split into chiral ``light" and ``mirror"
sectors, with strong ``mirror'' couplings to a unitary scalar in order to, hopefully,
lift the mirror sector.  Correlation functions of the obvious operators did not
show the presence of massless modes in the mirror sector.  However, because
the mirror spectrum is was anomalous, 't Hooft anomaly matching implied that they
ought to be present.  The follow-up work \cite{Poppitz:2009gt} employed the polarization tensor,
finding a directional discontinuity at zero momentum, a clear indication of
a massless state. 

Therefore, it is of interest to study a theory where 't Hooft anomaly matching does
not imply massless modes in the mirror sector.  Our recent work considered just such
a case \cite{Chen:2012di}, and we summarize our results in this proceedings report.
It will be seen that the massless modes persist.

\section{The 3-4-5 model}
\label{descr}
The ``3-4-5 model'' is a two-dimensional lattice gauge theory with
U(1) gauge invariance.  It has three Weyl fermion fields, which we denote $A_+$, $B_+$, $C_-$, 
with charge and chirality $3_+$, $4_+$, and $5_-$ respectively.  In addition, there
is a mirror sector, $3_-$, $4_-$, and $5_+$, and we call the 
respective fields $A_-$, $B_-$, and $C_+$.  In order to
construct all of the Yukawa couplings that are needed, a neutral spectator 
fermion $X_-$, with charge and chirality $0_-$, is also introduced,
together with its mirror sector partner $X_+$, with charge and 
chirality $0_+$.  Finally, in order to lift the
mirror sector, a unitary ``Higgs'' field $\phi$ with charge $-1$ is introduced.
The 3-4-5 model fields are given in Table \ref{fc}.

\begin{table}
\begin{center}
\begin{tabular}{|c|c|c|} \hline
Light Field & Mirror Field & Q \\ \hline
$A_+$ & $A_-$ & 3 \\
$B_+$ & $B_-$ & 4 \\
$C_-$ & $C_+$ & 5 \\
$X_-$ & $X_+$ & 0 \\
--- & $\phi$ & -1 \\ 
\hline
\end{tabular}
\end{center}
\caption{Summary of the field content in the 3-4-5 model. \label{fc}}
\end{table}

The dynamics of the gauge field is not supposed to be involved in the
mechanism that lifts the mirror fermions, so in our analysis we 
neglected the gauge field fluctuations and treated it only as a background field.
The action of the ``3-4-5'' model is
\beq 
\label{eq:action}
S &=& S_{\rm light}+S_{\rm mirror}    \nonumber  \\
S_{\rm light} &=& -(\bar{A}_+\cdot D_3\cdot A_+)
-(\bar{B}_+\cdot D_4\cdot B_+)-(\bar{C}_-\cdot D_5\cdot C_-)-(\bar{X}_-\cdot D_0\cdot X_-) \nonumber \\
S_{\rm mirror} &=&S_{\kappa}-(\bar{A}_-\cdot D_3\cdot A_-)
-(\bar{B}_- \cdot D_4\cdot B_-)-(\bar{C}_+ \cdot D_5\cdot C_+)-(\bar{X}_+ \cdot D_0\cdot X_+) \nonumber \\
           &&+S_{\text{Yuk.,Dirac}}+S_{\text{Yuk.,Maj}},
\eeq
Here, $D_q$ is Neuberger's overlap Dirac operator \cite{Neuberger:1997fp}, 
with charge $q$ on the gauge field.  Chiral projections are based on the $\gamma_5$ and
$\hat \gamma_5$ operators, as is typical in the overlap formalism.
The Yukawa interactions of the mirror sector are
\beq 
S_{\text{Yuk.,Dirac}} &=&
y_{30} \Abar_- X_+ \phi^{-3} + y_{40} \Bbar_- X_+ \phi^{-4} + y_{35} \Abar_- C_+ \phi^{2}
+ y_{45} \Bbar_- C_+ \phi \ddd + y_{30} \bar X_+ A_- \phi^3 + y_{40} \bar X_+ B_- \phi^4
+ y_{35} \bar C_+ A_- \phi^{-2} + y_{45} \bar C_+ B_- \phi^{-1}      \nonumber    \\ 
S_{\text{Yuk.,Maj.}} &=&
h_{30} A_-^T \gamma_2 X_+ \phi^3 + h_{40} B_-^T \gamma_2 X_+ \phi^4 
+ h_{35} A_-^T \gamma_2 C_+ \phi^8 + h_{45} B_-^T \gamma_2 C_+ \phi^9
\ddd - h_{30} \Xbar_+ \gamma_2 \Abar_-^T \phi^{-3} - h_{40} \Xbar_+ \gamma_2 \Bbar_-^T \phi^{-4}
- h_{35} \Cbar_+ \gamma_2 \Abar_-^T \phi^{-8} 
\ddd - h_{45} \Cbar_+ \gamma_2 \Bbar_-^T \phi^{-9}.
\label{eq:yukawa_interaction} 
\eeq
Each of the  terms in (\ref{eq:yukawa_interaction}) has an implicit sum 
over lattice sites:  $\Abar_- X_+ \phi^{-3} = \sum_x \Abar_{-,x} X_{+,x} \phi_x^{-3}$.
Finally, the field $\phi_x=e^{i \eta_x}$, $\left| \eta \right| \leq \pi$, 
is a unitary Higgs field of charge $-1$ with a kinetic term:
\beq
S_\kappa=\frac{\kappa}{2} \sum_x \sum_{\mu}[2-(\phi^*_x U^*_\mu(x) \phi_{x+\hat\mu}+h.c.) ]~.
\label{Sk}
\eeq

\section{Preliminaries of the calculation}
\subsection{The mirror polarization tensor}
The vacuum polarization tensor is defined as:
\beq \label{eq:Pi}
\left.\Pi_{\mu\nu}(x,y) \equiv \frac{\delta^2 \ln Z[A]}{\delta A(x) \delta A(y)} \right|_{A=0}.
\eeq
Just as the partition function  splits into the product of ``mirror" and ``light" parts, 
so does the polarization operator:
\beq 
\Pi_{\mu\nu}(x,y)=\Pi^{\text{light}}_{\mu\nu}(x,y) + \Pi^{\text{mirror}}_{\mu\nu}(x,y)~,
\label{eq:Pisplit}
\eeq

\subsection{Complex phase}
\label{cph}
The fermion determinant of our theory is not real.  
To overcome this feature, we have studied the dependence of the phase
distribution on the values of the Yukawa coupling parameters.  We have optimized
these parameters in order to have a narrow phase distribution, 
leading to the choice of couplings in Table \ref{tb:coupling1}.
Since we calculate the fermion determinant
at each sampling step of our simulation, we include the complex phase
in our expectation values.

\begin{table}
\centering
\begin{tabular}{cccccccc}
\hline
$h_{30}$ & $h_{40}$& $h_{35}$& $h_{45}$& $y_{30}$&$y_{40}$ & $y_{35}$& $y_{45} $ \\
\hline
   30.3214 & 3.08123 & 3.00278 & 23.7109 & 1.0 & 1.0 & 1.0 & 1.0  \\
\hline
\end{tabular}
\caption{A set of Yukawa coupling constants that we found by
optimizing narrowness of the complex phase distribution for an $8 \times 8$ lattice.  
\label{tb:coupling1}}
\end{table}

\subsection{Probing for massless particles}
In the continuum, the contribution to the Fourier transform 
of the real part of the polarization tensors due to 
massless particles takes the form:
\beq 
\label{eq:RePi}
\tilde\Pi_{\mu\nu}(k)= 2C \frac{\delta_{\mu\nu}k^2-k_\mu k_\nu}{k^2}~.
\eeq 
A single free charge-$q$ Weyl fermion, gives a contribution equal to 
that of a half charge-$q$ Dirac fermion:
$2C_{\text{fermion}} \simeq -\frac{1}{2\pi} q^2$.
The way to look for a  massless pole in (\ref{eq:RePi}) is to  notice that (\ref{eq:RePi}) 
has a directional limit as $k \rightarrow 0$:
\beq \label{eq:direclimit}
\left.\tilde\Pi_{11}(\phi) \right|_{k\rightarrow 0} = C(1-\cos 2\phi),  \quad
\left.\tilde\Pi_{21}(\phi) \right|_{k\rightarrow 0} = -C \sin 2\phi,
\eeq
where $\phi$ is the angle of approach to the origin measured from the positive-$k_1$ axis. 
Therefore if there is a massless particle in the spectrum
of the mirror theory, we would expect $\tilde\Pi_{\mu\nu}^{\text{mirror},\prime}$ 
has the following behavior as $k\rightarrow 0$:
\beq \label{eq:direclimit2}
\tilde\Pi_{11}(45^o)&=&-\tilde\Pi_{21}(45^o)=C~,   \quad
\tilde\Pi_{11}(90^o) = 2C~, \nonumber \\
\tilde\Pi_{11}(0^o)&=&\tilde\Pi_{21}(0^o)=\tilde\Pi_{21}(90^o)~.   
\eeq

On the other hand, if there is no massless particle, we would expect:
$\tilde\Pi_{\mu\nu} \sim (\delta_{\mu\nu} k^2-k_\mu k_\nu)/m^2$,
as $k\rightarrow 0$.  Therefore if the mirror sector acquires
the desired mass scale, the directional 
limit behavior (\ref{eq:direclimit}) disappears. 

\section{Results}
For the set of couplings given in Table \ref{tb:coupling1}, we
computed $\tilde\Pi_{\mu\nu}^{\text{mirror},\prime}(k)$ for
$6 \times 6$, $8 \times 8$ and $10 \times 10$ lattices.  
The reason for using such small lattices is because the calculation of 
$\tilde\Pi_{\mu\nu}^{\text{mirror},\prime}(k)$ 
is rather demanding and scales badly ($\sim N^{10}$ after
having taken into account momentum conservation) with
the number of lattice sites $N$ in each direction.
We studied three directions in momentum space:
$0^o$ with ${\bf k}=(k,0)$; $45^o$ with ${\bf k}=(k,k)$;
$90^o$ with ${\bf k}=(0,k)$.  Here $k$ was taken
from $0,1,\ldots,N-1$ on an $N \times N$ lattice.
The corresponding momenta in physical units are $p=2 \pi k/Na$.

\begin{figure}
\begin{minipage}{3.5in}
\includegraphics[width=3in,height=2in]{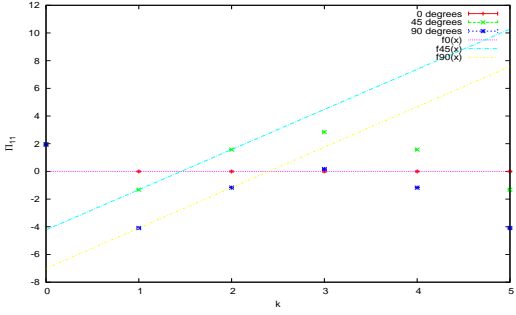}
\end{minipage}
\hfill
\begin{minipage}{2.5in}
\caption{$\tilde\Pi_{11}^{\text{mirror},\prime}(k)$ 
on a $6\times 6$ lattice.  The lines
show the extrapolation $k \to 0$ for different
angles of approach.  A clear discontinuity in
the directional limit can be seen.  \label{fig:Ne6}}
\end{minipage}
\end{figure}

Figure \ref{fig:Ne6} shows the results obtained on the $6 \times 6$ lattice.
It has the pattern suggested by (\ref{eq:direclimit}) with $C \approx 4 \approx 50/(4\pi)$.
In fact, it is remarkably similar to Figs.~2-5 of Ref.~\cite{Poppitz:2009gt},
except that the discontinuity is about 50 times larger, consistent
with contributions from the three charges that are present in the
underlying theory:  $3^2+4^2+5^2=50$.
The $k \to 0$ extrapolations are given in the second column of 
Table \ref{tb:fits}, where
we used a linear fit to the two smallest nonzero $k$ values.

\begin{table}
\begin{minipage}{3.5in}
\begin{tabular}{cccc}
\hline
 $\phi$ & $A (6 \times 6)$ & $A (8 \times 8)$ & $A (10 \times 10)$ \\
\hline
 $0^o$ & -$2.36 \times 10^{-3}$ & -0.16423(82) & --- (0)\\
 $45^o$ & -4.22 & -3.820(98) & -3.764(41) \\
 $90^o$ & -7.01 & -6.43(16) & -6.501(47) \\
\hline
\end{tabular}
\end{minipage}
\hfill
\begin{minipage}{2.5in}
\caption{Linear extrapolation $k \to 0$, using a fit to $f(k) = A + B k$.
The $0^o$ value for $N=10$ was not calculated because of computer time constraints;  
based on the results from $N=6,8$, as well as calculations
in the free theory, we expect the $0^o$ values to vanish, within errors.
\label{tb:fits}}
\end{minipage}
\end{table}

Our results for the $8\times 8$ and $10 \times 10$ lattices with
the same Yukawa couplings are similar to the $6 \times 6$ case, are shown in \cite{Chen:2012di}, 
and have
$k \to 0$ extrapolations given in the other
columns of Table \ref{tb:fits}.  As one can see, the discontinuity 
constant $C$ is approximately the same as was found on the $6 \times 6$ lattice.
 
\begin{figure} 
\begin{minipage}{3.5in}
\includegraphics[width=3in,height=2in]{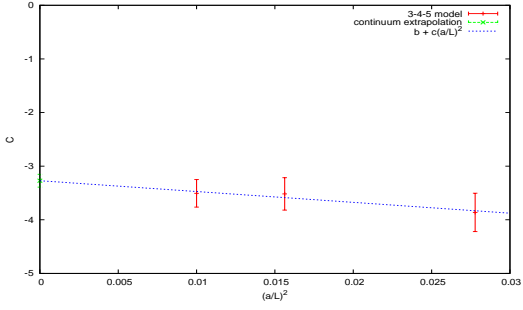}
\end{minipage}
\hfill
\begin{minipage}{2.5in}
\caption{The small-$k$ discontinuity in $\tilde{\Pi}_{11}$ fitted to a linear function of $(a/L)^2$.
It is seen that the discontinuity appears to approach a nonzero value as $a \to 0$. 
\label{fig:345discfit}}
\end{minipage}
\end{figure}

If the physical volume $L \times L$ is held fixed, where $L=Na$, while the number
of lattice sites $N \times N$ is increased, this corresponds to decreasing
the lattice spacing.  The discretization error of the overlap fermions which we are
using is known to be $\ord{a^2}$.  Thus continuum extrapolation obeys
\beq
C = b + c (a/L)^2 + \ord{(a/L)^4} = b + c N^{-2} + \ord{N^{-4}}
\eeq
Here, we use the values of $C$ from Table~\ref{tb:fits} for each of
the three values of $N=L/a$ that we have computed.  The errors
in the value of $C$ are dominated by the difference between the
value obtained from $45^o$ versus $90^o$.  Taking this into account,
we obtain the fit shown in Figure~\ref{fig:345discfit}.  It can
be seen that the linear function in $(a/L)^2$ describes the
data very well.  Furthermore, the linear extrapolation
intersects the $a=0$ limit at a nonzero discontinuity, $b= -3.27(12)$.  
This suggests that in the continuum
limit of the 3-4-5 model, the polarization tensor has a directional
discontinuity, consistent with having massless modes.

\section{Conclusions}
We have studied the polarization operator of the 3-4-5 mirror 
theory in order to probe for the existence of massless mirror states. 
We find evidence that the polarization tensor has a directional 
discontinuity at $k=0$.  This supports the conclusion that the mirror sector
fermions do not decouple, but remain massless, in spite of the strong Yukawa dynamics.
It is strange that the strong Yukawa interaction does not give mass to
the mirror sector, given that 't Hooft anomaly matching does not
imply any massless modes in this anomaly free theory.  One goal of
future work is to understand what the mechanism is for giving massless
modes in the mirror sector in this lattice theory.

\section*{Acknowledgments}
C.C.~and J.G.~were supported in part by the Department of Energy,
Office of Science, Office of High Energy Physics, 
Grant No.~DE-FG02-08ER41575.  
E.P.~was supported in part by the National Science 
and Engineering Council of Canada (NSERC).
We gratefully acknowledge
the use of USQCD computing resources at Fermi National
Laboratory, under a Class C allocation.

\bibliography{lattice2013}
\bibliographystyle{JHEP}

\end{document}